\documentclass[12pt]{article}
\usepackage{amssymb}
\usepackage{amsmath}
\usepackage[unicode,bookmarks,bookmarksopen,bookmarksopenlevel=2,colorlinks,linkcolor=blue,citecolor=green]{hyperref}

\input{tcilatex}
\begin{document}

\title{\textbf{Three Dimensional Reductions of Four-Dimensional Quasilinear
Systems}}
\author{Maxim V. Pavlov$^{1,2,3}$, Nikola M. Stoilov$^{4,5}$ \\
$^{1}$Sector of Mathematical Physics,\\
Lebedev Physical Institute of Russian Academy of Sciences,\\
Leninskij Prospekt 53, 119991 Moscow, Russia\\
$^{2}$Department of Applied Mathematics,\\
National Research Nuclear University MEPHI,\\
Kashirskoe Shosse 31, 115409 Moscow, Russia,\\
$^{3}$Department of Mechanics and Mathematics, Novosibirsk State University,
\ \ \ \ \ \ \\
2 Pirogova street, Novosibirsk, 630090, Russia\\
$^{4}$Institut de Math\'ematiques de Bourgogne,\\
Universit\'e de Bourgogne, 9 avenue Alain Savary, 21078 Dijon Cedex, France.%
\\
$^{5}$Max Plank Institute for Dynamics and Self-Organisation,\\
37077, G\"{o}ttingen, Germany.}
\date{}
\maketitle

\begin{abstract}
In this paper we show that integrable four-dimensional linearly degenerate
equations of second order possess infinitely many three-dimensional
hydrodynamic reductions. Furthermore, they are equipped with infinitely many
conservation laws and higher commuting flows. We show that the dispersionless
limits of nonlocal KdV and nonlocal NLS equations (the so-called Breaking
Soliton equations introduced by O.I. Bogoyavlenski) are one and two
component reductions (respectively) of one of these four-dimensional
linearly degenerate equations.
\end{abstract}

\tableofcontents

keywords: hydrodynamic chain, hydrodynamic reduction, dispersionless limit,
breaking soliton.

\section{Introduction}

This work is inspired by a study of Bogoyavlenskii's Breaking Soliton
equations and especially their dispersionless limit. These equations
arise as a simple two-dimensional generalisation of well-known equations, by
allowing the Lax pair to depend on an additional independent variable. The
analogue of KdV, often called the Breaking Soliton equation (see \cite{B90}),
written in its nonlocal form is 
\begin{equation}
v_{t}-\frac{1}{2}v_{y}\partial^{-1}_yv_{x}-vv_{x}+\frac{\epsilon ^{2}}{2}%
v_{xyy}=0.  \label{kdv}
\end{equation}%
This equation is integrable,
possesses a Lax pair and infinitely many commuting flows. What is remarkable
is that its dispersionless limit 
\begin{equation}
v_{t}+vv_{x}+uv_{y}=0,\quad u_{y}=\frac{1}{2}v_{x},  \label{breakingdisp}
\end{equation}%
cannot be treated with the standard integrability test for multidimensional
quasilinear systems, based on the method of hydrodynamic reductions, since
the dispersion relation is degenerate - it reduces to two lines rather than
being a conic (for details see Section 2 and \cite{FK, ZhenjaKarima}).
Furthermore, the (2+1)-dimensional non-linear Schr\"{o}dinger equation,
which also appears in \cite{B90} as a breaking soliton generalisation of NLS 
\begin{equation}
i\epsilon \,\psi _{t}+\epsilon ^{2}\psi _{xy}\pm 2\,\psi \partial^{-1}_y(|\psi
|^{2})_{x}\,=0,\,  \label{nls}
\end{equation}%
after an appropriate transformation (the so called Madelung transformation,
see Section 5), gives rise in a dispersionless limit to 
\begin{equation}
R_{t}^{1}+R^{1}R_{x}^{1}+uR_{y}^{1}=0,\quad
R_{t}^{2}+R^{2}R_{x}^{2}+uR_{y}^{2}=0,\quad u_{y}=\frac{1}{2}%
(R^{1}+R^{2})_{x}.  \label{NLS2+1disp}
\end{equation}%
Since both nonlocal systems (\ref{kdv}) and (\ref{nls}) are integrable,
their dispersionless limits (\ref{breakingdisp}) and (\ref{NLS2+1disp}) are also
 integrable (because they preserve infinitely many conservation laws and
higher commuting flows). So the question of how to understand their
integrability, and more generally the integrability of the
generalisation to $M$ components ($\kappa _{i}$ are constants) 
\begin{equation*}
R_{t}^{i}+R^{i}R_{x}^{i}+uR_{y}^{i}=0,\quad u_{y}=\sum_{i=1}^{M}\kappa
_{i}R_{x}^{i},\quad i=1,\dots ,M,
\end{equation*}%
arises naturally\footnote{%
The first system (\ref{breakingdisp}) is linearisable by a point
transformation of the dependent and independent variable $x=x(v,t)$. Such an
approach, however, does not generalise to the multicomponent case.}.

As already mentioned, the method of hydrodynamic reductions provides a
standard, constructive test for the integrability of multidimensional
quasilinear systems of first order. A key point is that this method is based
on the existence of sufficiently many \emph{two-dimensional} hydrodynamic
reductions (see again \cite{FK}, \cite{ZhenjaKarima}).

In this paper we show that integrable linearly degenerate four-dimensional
equations of second order also possess infinitely many \emph{three-dimensional} hydrodynamic reductions.

Among the simplest examples of linearly degenerate four dimensional
integrable equations are (see, for instance, \cite{ZhenjaKarima}, \cite%
{Takasaki}, \cite{BogdMaks}):%
\begin{equation*}
U_{y\tau }=U_{xy}U_{z}-U_{y}U_{xz},
\end{equation*}%
\begin{equation*}
U_{x\tau }=U_{tz}+U_{xx}U_{z}-U_{x}U_{xz},
\end{equation*}%
\begin{equation}
U_{\sigma \tau }=U_{zz}+U_{z}U_{x\sigma }-U_{\sigma }U_{xz}.  \label{third}
\end{equation}%
In all these examples $U = U(x,t,y,z,\tau ,\sigma )$. These four-dimensional
quasilinear equations are determined by the following dispersionless Lax
pairs ( where $\lambda $ is an arbitrary parameter)%
\begin{equation*}
\psi _{y}=-\frac{1}{\lambda }U_{y}\psi _{x},\text{ \ \ }\psi _{\tau
}=\lambda \psi _{z}+U_{z}\psi _{x},
\end{equation*}%
\begin{equation*}
\psi _{t}=(\lambda +U_{x})\psi _{x},\text{ \ \ }\psi _{\tau }=\lambda \psi
_{z}+U_{z}\psi _{x},
\end{equation*}%
\begin{equation}
\psi _{z}=\lambda \psi _{\sigma }+U_{\sigma }\psi _{x},\text{ \ \ }\psi
_{\tau }=\lambda \psi _{z}+U_{z}\psi _{x},  \label{tri}
\end{equation}%
respectively, where $\psi = \psi (x,t,y,z,\tau ,\sigma ;\lambda )$ in all
cases.

The paper is organised as follows: in Section 2 we consider the method of
hydrodynamic reductions and its applicability to the aforementioned systems.
In Section 3 we introduce three-dimensional hydrodynamic chains arising from
these systems. In Section 4 we introduce multidimensional hydrodynamic
reductions. We return to Breaking Soliton equations in Section 5.

\section{The Method of Hydrodynamic Reductions}

Without loss of generality we consider the third quasilinear equation (\ref%
{third}) from the previous Section (here we only changed the independent variables)%
\begin{equation}
U_{x}U_{ty}+U_{xz}=U_{t}U_{xy}+U_{tt}.  \label{w}
\end{equation}%
Introducing new variables  such that $u=U_{x}$ and $
a=U_{t}$) we obtain the four-dimensional two component quasilinear system%
\begin{equation}
u_{t}=a_{x},\text{ \ }ua_{y}+u_{z}=au_{y}+a_{t}  \label{quasidva}
\end{equation}%
determined by the dispersionless Lax pair (see (\ref{tri}))%
\begin{equation}
\psi _{t}=\lambda \psi _{x}+u\psi _{y},\text{ \ }\psi _{z}=\lambda \psi
_{t}+a\psi _{y}.  \label{laxpara}
\end{equation}%
This dispersionless Lax pair is a reduction of the more general
dispersionless Lax pair%
\begin{equation*}
\psi _{t}=(\lambda +v)\psi _{x}+u\psi _{y},\text{ \ }\psi _{z}=(\lambda
+p)\psi _{t}+a\psi _{y},
\end{equation*}%
which belongs to the class of hyper-K\"{a}hler hierarchies (see detail, for
instance, in \cite{Takasaki}). Moreover, the dispersionless Lax pair (\ref%
{laxpara}) together with the quasilinear system (\ref{quasidva}) is a four-
dimensional reduction (such that $s=-t$ and $\partial _{r}=0$) of the six-dimensional 
two-component quasilinear system (see \cite{ZhenjaKarima})%
\begin{equation*}
u_{t}=ua_{r}-au_{r}-a_{x},\text{ \ }a_{s}=au_{y}-ua_{y}+u_{z},
\end{equation*}%
determined by the dispersionless Lax pair%
\begin{equation*}
\psi _{s}+\lambda \psi_{x}+u\psi _{y}-\lambda u\psi _{r}=0,\text{ \ }\psi
_{z}-\lambda \psi_{t}+a\psi _{y}-\lambda a\psi _{r}=0.
\end{equation*}

In order to apply the method of hydrodynamic reductions (see detail in \cite%
{ZhenjaKarima}) we are looking for two-dimensional reductions in the form%
\begin{equation}
r_{t}^{i}=\mu ^{i}r_{x}^{i},\text{ \ }r_{y}^{i}=\zeta ^{i}r_{x}^{i},\text{ \ 
}r_{z}^{i}=\eta ^{i}r_{x}^{i},\text{ \ }i=1,2,...N,  \label{c}
\end{equation}%
where $N$ is an arbitrary natural number. This means that $u(x,t,y,z)=\tilde{%
u}(\mathbf{r}(x,t,y,z))$, $a(x,t,y,z)=\tilde{a}(\mathbf{r}(x,t,y,z))$, where
the $N$ Riemann invariants $r^{i}(x,t,y,z)$ simultaneously solve the three
commuting systems (\ref{c}).

We obtain 
two consequences, the so called dispersion relation%
\begin{equation}
\eta ^{i}=\tilde{a}\zeta ^{i}-\tilde{u}\zeta ^{i}\mu ^{i}+(\mu ^{i})^{2}
\label{relation}
\end{equation}%
(as usual here: $\partial _{i}\equiv \partial /\partial r^{i}$) and the relation between two conservation law densities $\tilde{u},\tilde{a}$
and the characteristic velocity $\mu ^{k}$%
\begin{equation}
\mu ^{i}\partial _{i}\tilde{u}=\partial _{i}\tilde{a}.  \label{sec}
\end{equation}%
Taking into account the Tsarev conditions\footnote{%
The Tsarev conditions follow from the compatibility conditions $%
(r_{t}^{i})_{y}=(r_{y}^{i})_{t},(r_{t}^{i})_{z}=(r_{z}^{i})_{t},(r_{z}^{i})_{y}=(r_{y}^{i})_{z} 
$, where $N$ Riemann invariants $r^{i}(x,t,y,z)$ are common unknown
functions for all three commuting flows (\ref{c}).} (see detail in \cite%
{Tsar})%
\begin{equation*}
\frac{\partial _{k}\mu ^{i}}{\mu ^{k}-\mu ^{i}}=\frac{\partial _{k}\zeta ^{i}%
}{\zeta ^{k}-\zeta ^{i}}=\frac{\partial _{k}\eta ^{i}}{\eta ^{k}-\eta ^{i}}
\end{equation*}%
and verifying the compatibility conditions $\partial _{k}(\partial _{i}%
\tilde{a})=\partial _{i}(\partial _{k}\tilde{a})$, we obtain the
Gibbons-Tsarev type system (cf. \cite{gibtsar}, \cite{ZhenjaKarima})%
\begin{equation}
\partial _{k}\zeta ^{i}=\frac{\zeta ^{i}(\zeta ^{k}-\zeta ^{i})}{\mu
^{k}-\mu ^{i}-\tilde{u}(\zeta ^{k}-\zeta ^{i})}\partial _{k}\tilde{u},\text{
\ \ }\partial _{k}\mu ^{i}=\frac{\zeta ^{i}(\mu ^{k}-\mu ^{i})}{\mu ^{k}-\mu
^{i}-\tilde{u}(\zeta ^{k}-\zeta ^{i})}\partial _{k}\tilde{u},  \label{gt}
\end{equation}%
\begin{equation}
\partial _{ik}\tilde{u}=\frac{\zeta ^{i}-\zeta ^{k}}{\mu ^{k}-\mu ^{i}-%
\tilde{u}(\zeta ^{k}-\zeta ^{i})}\partial _{i}\tilde{u}\partial _{k}\tilde{u}%
.  \label{uk}
\end{equation}%
This system is in involution. Any particular solution (a general solution
depends on $2N$ arbitrary functions of a single variable) determines three
commuting hydrodynamic type systems (\ref{quasidva}), which can be
integrated by the Tsarev generalised hodograph method (see detail in \cite%
{Tsar}). Each of these hydrodynamic type systems possesses a general
solution parameterised by $N$ arbitrary functions of a single variable.
Thus, the method of two dimensional hydrodynamic reductions yields solutions
parameterised by $3N$ arbitrary functions of a single variable.

Now we introduce the auxiliary function $b$ such that $b=U_{y}$. This means
that $b_{t}=a_{y}$ and $b_{x}=u_{y}$. Then the corresponding function $%
\tilde{b}(\mathbf{r}(x,t,y,z))=b(x,t,y,z)$ satisfies the relationship
between two conservation law densities $\tilde{b},\tilde{u}$ and the
characteristic velocity $\zeta ^{k}$%
\begin{equation}
\partial _{i}\tilde{u}=\frac{1}{\zeta ^{i}}\partial _{i}\tilde{b}.
\label{tre}
\end{equation}%
Then equations (\ref{uk}) reduce to the form%
\begin{equation*}
\partial _{ik}\tilde{b}=0.
\end{equation*}%
Hense, up to reparametrisations $r^{i}\rightarrow \varphi _{i}(r^{i})$, one
has\footnote{%
See similar computations in \cite{ZhenjaKarima}, last formulas on page 2371.}%
\begin{equation*}
\tilde{b}=\sum_{m=1}^{N}r^{m}.
\end{equation*}%
Then other equations (\ref{relation}), (\ref{gt}) become (here $f_{i}(r^{i})$
are arbitrary functions)%
\begin{equation*}
\zeta ^{i}=\frac{1}{v^{i}},\text{ \ }\mu ^{i}=f_{i}(r^{i})+\frac{\tilde{u}}{%
v^{i}},\text{ \ }\eta ^{i}=f_{i}^{2}(r^{i})+f_{i}(r^{i})\frac{\tilde{u}}{%
v^{i}}+\frac{\tilde{a}}{v^{i}},
\end{equation*}%
where \footnote{%
See again similar computations in \cite{ZhenjaKarima}, the first formula on
page 2372. The integrability of system (\ref{lin}) is presented in \cite%
{maksint}.}%
\begin{equation}
\frac{\partial _{k}v^{i}}{v^{k}-v^{i}}=\frac{1}{f_{k}(r^{k})-f_{i}(r^{i})}.
\label{lin}
\end{equation}%
Thus the Gibbons-Tsarev type system (\ref{gt})-(\ref{uk}) determines the commuting triple of
two dimensional hydrodynamic type systems (\ref{c}), where the functions $%
\tilde{u}$ and $\tilde{a}$ can be found by quadratures (see (\ref{tre}) and (%
\ref{sec}), respectively):%
\begin{equation*}
d\tilde{u}=\sum_{m=1}^{N}v^{m}dr^{m},\text{ \ }d\tilde{a}%
=\sum_{m=1}^{N}(f_{m}(r^{m})v^{m}+\tilde{u})dr^{m}.
\end{equation*}%
The integrability of these hydrodynamic type systems (\ref{c}) was investigated
in \cite{maksint}. Thus, the integrability (by the method of two dimensional
hydrodynamic reductions) of the four-dimensional quasilinear equation of
second order (\ref{quasidva}) is reduced to a construction of the general
solution for the hydrodynamic type systems (\ref{c}).

In the next Section we present a three-dimensional hydrodynamic chain associated
with the quasilinear system (\ref{quasidva}) and its dispersionless Lax pair
(\ref{laxpara}). Such three-dimensional hydrodynamic chains are a convenient
tool for the construction of three dimensional hydrodynamic reductions.

\section{Three-Dimensional Hydrodynamic Chains}

Under the potential substitution $h=\psi _{y}$ the dispersionless Lax pair (%
\ref{laxpara}) takes the form%
\begin{equation*}
h_{t}=\lambda h_{x}+(uh)_{y},\text{ \ }h_{z}=\lambda ^{2}h_{x}+[(\lambda
u+a)h]_{y}.
\end{equation*}%
The asymptotic expansion at ($\lambda \rightarrow \infty $)%
\begin{equation*}
h=\exp \left( -\overset{\infty }{\underset{k=0}{\sum }}\frac{A^{k}}{\lambda
^{k+1}}\right) =1-\frac{h_{0}}{\lambda }-\frac{h_{1}}{\lambda ^{2}}-\frac{%
h_{2}}{\lambda ^{3}}-...
\end{equation*}%
leads to a pair of commuting three-dimensional hydrodynamic chains,%
\begin{equation}
A_{t}^{k}=A_{x}^{k+1}+uA_{y}^{k},\text{ \ }k=0,1,...,  \label{ak}
\end{equation}%
\begin{equation}
A_{z}^{k}=A_{x}^{k+2}+uA_{y}^{k+1}+aA_{y}^{k},\text{ \ }k=0,1,...,  \label{b}
\end{equation}%
with two constraints:%
\begin{equation}
u_{y}=A_{x}^{0},\text{ \ }a_{y}=A_{t}^{0}.  \label{s}
\end{equation}%
These hydrodynamic chains possess infinitely many conservation laws:%
\begin{equation}
(h_{k})_{t}=(h_{k+1})_{x}+(uh_{k})_{y},\text{ }k=0,1,...,  \label{conserv}
\end{equation}%
\begin{equation*}
(h_{k})_{z}=(h_{k+2})_{x}+(uh_{k+1}+ah_{k})_{y},
\end{equation*}%
where the first three conservation law densities are%
\begin{equation*}
h_{0}=A^{0},\text{ \ }h_{1}=A^{1}-\frac{1}{2}(A^{0})^{2},\text{ \ }%
h_{2}=A^{2}-A^{0}A^{1}+\frac{1}{6}(A^{0})^{3}.
\end{equation*}%
The two constraints (\ref{s}) reduce to two additional conservation laws%
\begin{equation*}
u_{y}=(h_{0})_{x},\text{ \ }(a-uh_{0})_{y}=(h_{1})_{x}.
\end{equation*}

\textbf{Remark}: The quasilinear equation (\ref{w}) can be derived from the
hydrodynamic chains (\ref{ak}) and (\ref{b}), extracting the first two
equations from (\ref{ak}), zeroth equation from (\ref{b}) and both
constraints (\ref{s}), i.e.%
\begin{equation*}
A_{t}^{0}=A_{x}^{1}+uA_{y}^{0},\text{ \ \ }A_{t}^{1}=A_{x}^{2}+uA_{y}^{1},%
\text{ \ }A_{z}^{0}=A_{x}^{2}+uA_{y}^{1}+aA_{y}^{0},\text{ \ }%
u_{y}=A_{x}^{0},\text{ \ }a_{y}=A_{t}^{0}.
\end{equation*}%
Eliminating $A_{x}^{2}$, 
\begin{equation}
A_{t}^{0}=A_{x}^{1}+uA_{y}^{0},\text{ \ }A_{z}^{0}=A_{t}^{1}+aA_{y}^{0},%
\text{ \ \ }u_{y}=A_{x}^{0},\text{ \ }a_{y}=A_{t}^{0},  \label{4}
\end{equation}%
where%
\begin{equation*}
A_{x}^{2}=A_{t}^{1}-uA_{y}^{1}.
\end{equation*}%
Introducing a potential function $U$ such that $u=U_{x},a=U_{t}$ and $%
A^{0}=U_{y}$, (\ref{4}) reduces to the following pair of equations:%
\begin{equation*}
A_{x}^{1}=U_{yt}-U_{x}U_{yy},\text{ \ }A_{t}^{1}=U_{yz}-U_{t}U_{yy},
\end{equation*}%
with compatibility condition $(A_{x}^{1})_{t}=(A_{t}^{1})_{x}$ leading to 
\begin{equation*}
U_{ytt}-U_{x}U_{yyt}=U_{xyz}-U_{t}U_{xyy},
\end{equation*}%
which is nothing but the derivative (with respect to independent variable $y$%
) of (\ref{w}).

\section{Three-Dimensional Hydrodynamic Reductions}

In this Section we extract the most natural three-dimensional hydrodynamic
reductions, i.e. $M$ component three-dimensional quasilinear systems, where $
M$ is an arbitrary natural number.

\textbf{I}. The first reduction is given by the constraint $A^{M}=$ const.
Then (\ref{ak}) reduces to the following multi-component three dimensional hydrodynamic
type systems%
\begin{equation*}
u_{y}=A_{x}^{0},\text{ \ }A_{t}^{M-1}=uA_{y}^{M-1},\text{ \ }%
A_{t}^{k}=A_{x}^{k+1}+uA_{y}^{k},\text{ \ }k=0,1,...,M-2,
\end{equation*}%
where $A^{M-1}$ can be recognised as a Riemann invariant. For
instance, if $M=1$, then%
\begin{equation*}
u_{y}=A_{x}^{0},\text{ \ }A_{t}^{0}=uA_{y}^{0};
\end{equation*}%
if $M=2$, then%
\begin{equation*}
u_{y}=A_{x}^{0},\text{ \ }A_{t}^{0}=A_{x}^{1}+uA_{y}^{0},\text{ \ }%
A_{t}^{1}=uA_{y}^{1};
\end{equation*}%
if $M=3$, then%
\begin{equation*}
u_{y}=A_{x}^{0},\text{ \ }A_{t}^{0}=A_{x}^{1}+uA_{y}^{0},\text{ \ }%
A_{t}^{1}=A_{x}^{2}+uA_{y}^{1},\text{ \ }A_{t}^{2}=uA_{y}^{2},
\end{equation*}%
etc.

\textbf{II}. The second reduction is given by the constraint $h_{M}=$ const.
Then (\ref{conserv}) reduces to a set of multi-component three-dimensional
hydrodynamic type systems%
\begin{equation*}
u_{y}=(h_{0})_{x},\text{ \ }(h_{N-1})_{t}=(uh_{N-1})_{y},\text{ \ }%
(h_{k})_{t}=(h_{k+1})_{x}+(uh_{k})_{y},\text{ }k=0,1,...,M-2.
\end{equation*}%
For instance, if $M=1$, then%
\begin{equation*}
u_{y}=(h_{0})_{x},\text{ \ }(h_{0})_{t}=(uh_{0})_{y};
\end{equation*}%
if $M=2$, then%
\begin{equation*}
u_{y}=(h_{0})_{x},\text{ \ }(h_{0})_{t}=(h_{1})_{x}+(uh_{0})_{y},\text{ \ }%
(h_{1})_{t}=(uh_{1})_{y},
\end{equation*}%
if $M=3$, then%
\begin{equation*}
u_{y}=(h_{0})_{x},\text{ \ }(h_{0})_{t}=(h_{1})_{x}+(uh_{0})_{y},\text{ \ }%
(h_{1})_{t}=(h_{2})_{x}+(uh_{1})_{y},\text{ \ }(h_{2})_{t}=(uh_{2})_{y},
\end{equation*}%
and so on.

\textbf{III}. The ansatz\footnote{%
This is the so-called \textquotedblleft waterbag\textquotedblright\
reduction; see, for instance, \cite{maksbenney}} ($\kappa _{i}$ are
arbitrary constants)%
\begin{equation*}
A^{k}=\frac{1}{k+1}\overset{M}{\underset{i=1}{\sum }}\kappa
_{i}(R^{i})^{k+1},\text{ }k=0,1,...
\end{equation*}%
reduces (\ref{ak}) and (\ref{b}) to two commuting $M$-component three-
dimensional hydrodynamic type systems%
\begin{equation}
R_{t}^{i}=R^{i}R_{x}^{i}+uR_{y}^{i},\text{ \ }%
R_{z}^{i}=(R^{i})^{2}R_{x}^{i}+(uR^{i}+a)R_{y}^{i},  \label{a}
\end{equation}%
where%
\begin{equation}
u_{y}=\left( \overset{M}{\underset{m=1}{\sum }}\kappa _{m}R^{m}\right) _{x},%
\text{ \ }a_{y}=\left( \overset{M}{\underset{m=1}{\sum }}\kappa
_{m}R^{m}\right) _{t}.  \label{e}
\end{equation}

\textbf{Lemma}: \textit{The two} $M$ \textit{component three-dimensional
hydrodynamic type systems} (\ref{a}) and (\ref{e}) \textit{commute with each
other if and only if the functions} $u(x,t,y,z),a(x,t,y,z)$ \textit{satisfy}
(\ref{quasidva}).

\textbf{Proof}: Given by a straightforward computation.

This $M$ parametric family of $M$ component three-dimensional hydrodynamic
reductions can be generalised to a family of $M$ component three-dimensional
hydrodynamic reductions parameterised by $M$ arbitrary functions of a single
variable.

\textbf{Theorem}: \textit{The four-dimensional quasilinear system }(\ref%
{quasidva}) \textit{possesses }$M$ \textit{component hydrodynamic reductions 
}(\ref{a}), (\ref{e})\textit{\ where the functions} $u(x,t,y,z)$, $%
a(x,t,y,z) $ \textit{are determined by}%
\begin{equation*}
u_{y}=A_{x}^{0},\text{ \ }a_{y}=A_{t}^{0},
\end{equation*}%
\textit{and all moments }$A^{k}(\mathbf{R})$ \textit{are parameterised by }$%
M $ \textit{arbitrary functions }$f_{0k}(R^{k})$ \textit{of a single variable%
}%
\begin{equation}
A^{k}=\overset{M}{\underset{m=1}{\sum }}f_{km}(R^{m}),  \label{mom}
\end{equation}%
\textit{where }$f_{k+1,i}^{\prime }(R^{i})=R^{i}f_{ki}^{\prime }(R^{i})$, $%
k=0,1$, \dots

\textbf{Proof}: Substituting $A^{k}(\mathbf{R})$ into (\ref{ak}) and (\ref{b}%
) leads to a sole relationship (here $\partial _{i}\equiv \partial /\partial
R^{i}$)%
\begin{equation}
\partial _{i}A^{k+1}=R^{i}\partial _{i}A^{k}.  \label{constr}
\end{equation}%
Compatibility conditions $\partial _{j}(\partial _{i}A^{k+1})=\partial
_{i}(\partial _{j}A^{k+1})$ yield (\ref{ak}). Substituting (\ref{mom}) into (%
\ref{constr}) implies $f_{k+1,i}^{\prime }(R^{i})=R^{i}f_{ki}^{\prime
}(R^{i})$. The Theorem is proved.\newline

All three-dimensional hydrodynamic reductions discussed above should be
considered as integrable quasilinear systems, because they possess
infinitely many conservations laws (see (\ref{conserv})). Since we deal with
hyper-K\"{a}hler hierarchies, all higher commuting flows are known (see
detail, for instance, in \cite{BogdMaks} and \cite{Takasaki}). For instance,
the next commuting three-dimensional hydrodynamic chain in the hierarchy is%
\begin{equation*}
A_{t^{3}}^{k}=A_{x}^{k+3}+uA_{y}^{k+2}+aA_{y}^{k+1}+cA_{y}^{k},
\end{equation*}%
where%
\begin{equation*}
u_{y}=A_{x}^{0},\text{ \ }a_{y}=A_{t}^{0},\text{ \ }c_{y}=A_{z}^{0}.
\end{equation*}%
%
%
%
%
In our opinion, the existence of infinitely many higher commuting flows is
sufficient for aforementioned reductions to be integrable. The problem of
obtaining solutions is open and will be discussed elsewhere.

\section{Breaking Soliton Equations}

In the one-component case equations (\ref{a}) and (\ref{e}) with $\kappa
=1/2 $, read: 
\begin{equation*}
R_{t}=RR_{x}+uR_{y},\quad R_{z}=(R)^{2}R_{x}+(uR+a)R_{y},
\end{equation*}%
where%
\begin{equation*}
u_{y}=\frac{1}{2}R_{x},\quad a_{y}=\frac{1}{2}R_{t}.
\end{equation*}%
Then the $z$ equation above is nothing but the dispersionless limit of the
second member of Bogojavlenskii's breaking soliton hierarchy \footnote{%
The entire hierarchy can be constructed by consecutively applying the KdV
recursion operator $\mathcal{R} =-\frac{\epsilon}{2}\partial_y^2 + v + \frac{1}{2} v_y\partial^{-1}_y$ to $u_{x}$} 
\begin{equation*}
\begin{array}{c}
v_{z}=v^{2}v_{x}+\frac{3}{2}(\partial^{-1}_y v_{x})vv_{y}+\frac{1}{4}%
v_{y}\partial^{-1}_y(v^{2})_{x} \\ 
~ \\ 
+\epsilon ^{2}\left[ -2(vv_{yy})_{x}+\frac{3}{4}\left( (\partial^{-1}_y v_{x})v_{x}\right) _{xx}-(\partial^{-1}_y v_{x})v_{yyy}-\frac{1}{2}%
(v_{x}v_{y})_{y}-\frac{1}{6}v_{xyy}-\frac{1}{12}v_{y}v_{xy}\right] +\epsilon
^{4}\frac{1}{8}v_{xyyyy}.%
\end{array}%
\end{equation*}%
The dispersionless limit is 
\begin{equation*}
v_{z}=v^{2}v_{x}+\frac{1}{4}(3v\partial^{-1}_y v_{x}+\partial
^{-1}_y(v^{2})_{x})v_{y}.
\end{equation*}%
Taking into account that $\partial^{-1}_y vv_{x}=4a-2vu$, we arrive exactly at $%
v_{z}=(v)^{2}v_{x}+(uv+a)v_{y}$. So we have shown that first two commuting flows
of the nonlocal KdV hierarchy in the dispersionless limit are three-dimensional one-component 
reductions of the four-dimensional linearly degenerate
equation of second order (\ref{quasidva}).

The two-component case is, in a similar way, related to the higher flows of
the (2+1)-dimensional NLS. The second flow is 
\begin{equation}
\psi _{z}=\epsilon ^{2}\psi _{xyy}-4\psi \psi ^{\ast }\psi _{x}-2\psi
_{y}\partial^{-1}_y(\psi \psi ^{\ast })_{x}+2\psi \partial^{-1}_y(\psi _{y}\psi _{x}^{\ast
}-\psi _{x}\psi _{y}^{\ast })  \label{NLS2}
\end{equation}%
In order to obtain the dispersionless limit and derive equation (\ref%
{NLS2+1disp}) in Section 1 we use a Madelung transformation $\psi =\sqrt{%
\rho }e^{iS/\epsilon }$ and introduce the Riemann invariants $R^{1}$ and $%
R^{2}$ by 
\begin{equation*}
R^{1}=S_{y}-2\sqrt{\rho },\quad R^{2}=S_{y}+2\sqrt{\rho },\quad S_{x}=u.
\end{equation*}%
We do the same here. Then the hydrodynamic limit of (\ref{NLS2}) becomes 
\begin{equation*}
R_{z}^{i}={R^{i}}^{2}R_{x}+(uR^{i}+a)R_{y},\quad u_{y}=\frac{1}{2}%
(R^{1}+R^{2})_{x},\quad a_{y}=\frac{1}{2}(R^{1}+R^{2})_{t}
\end{equation*}%
As in the normal case for multidimensional hydrodynamic systems, each higher
flow will introduce a \textquotedblleft higher\textquotedblright\ nested
non-locality, for instance the next one is $c_{y}=\sum_{i=1}^{N}\kappa
^{i}R_{z}^{i}$.

\section{Conclusion}

In this paper we considered the four-dimensional quasilinear system (\ref%
{quasidva}) determined by its dispersionless Lax pair (\ref{laxpara}). We
extracted $M$ component three-dimensional hydrodynamic reductions. Although we could not present a method for obtaining general solutions for these reductions, we believe that they are integrable because they possess infinitely many conservation laws and commuting flows. Moreover, our approach is universal, meaning that any $D$-dimensional quasilinear system from any hyper-K\"{a}hler hierarchy possesses 
$(D-1)$-dimensional hydrodynamic reductions.

Any multi-dimensional linearly degenerate equation of second order, which necessarily 
belongs to hyper-K\"{a}hler hierarchy possesses infinitely many global
solutions (see, for instance, \cite{FKK09}). However, the Breaking Soliton
equations have no global solutions. Thus the extraction of corresponding
solutions from multi-dimensional linearly degenerate equations of second
order is a fascinating and open problem.

\section*{Acknowledgements}

The authors thank Eugene Ferapontov for his involvement at the early stage of
this work, and numerous discussions during the process of completing the paper. NMS would like to thank Folkert
M\"uller-Hoissen for some very encouraging discussions.

MVP's work was partially supported by the grant of Presidium of RAS
\textquotedblleft Fundamental Problems of Nonlinear
Dynamics\textquotedblright\ and by the RFBR grant 14-01-00389.

NMS's work is partially supported by the Marie Curie-Sk\l odowska Intra-European
Fellowship for the  HYDRON project - ``Near Hydrodynamic type systems in 2+1
dimensions''. 

\end{document}